\documentclass[12pt]{article}
\usepackage{amssymb,amsmath} 
\usepackage{graphicx}
\graphicspath{{graphics/}}
\usepackage{latexsym}
\usepackage{subfig}  
\usepackage{booktabs,setspace,caption}  
\usepackage{braket}
\usepackage{wasysym,verbatim}
\usepackage[numbers,sort&compress]{natbib}
\usepackage{textcomp}
\usepackage[utf8]{inputenc}

\usepackage{mathrsfs}
\usepackage[inner=2.9cm,outer=2cm]{geometry}
\usepackage[toc]{appendix}
\usepackage{color,soul} 
\usepackage{datetime}
\usepackage[
      colorlinks=true,
      linkcolor=darkblue,  
      urlcolor=black,    
      filecolor=blue,     
      citecolor=blue, 
      linktocpage=true,
      pdfstartview=FitV,
      bookmarksopen=true    
      ]{hyperref}

\definecolor{darkblue}{rgb}{0.2, 0, 0.8}
\definecolor{darkgreen}{rgb}{0.2, 0.71, 0}

\newcommand{\eq}{\begin{equation}}
\newcommand{\qe}{\end{equation}}

\numberwithin{equation}{section}

\newenvironment{changemargin}[2]{%
\begin{list}{}{%
\setlength{\topsep}{0pt}%
\setlength{\leftmargin}{#1}%
\setlength{\rightmargin}{#2}%
\setlength{\listparindent}{\parindent}%
\setlength{\itemindent}{\parindent}%
\setlength{\parsep}{\parskip}%
}%
\item[]}{\end{list}}

\begin{document}  


\begin{titlepage}

\begin{flushright}
{\tt \small{IFUM-1077-FT}} \\
\end{flushright}

\vspace*{1.2cm}

\begin{center}
{\Large {\bf{Exact charges from heterotic black holes} }} \\

\vspace*{1.2cm}
\renewcommand{\thefootnote}{\alph{footnote}}
{\sl\large Federico Faedo$^{\text{\quarternote , \eighthnote}}$ and Pedro F.~Ram\'{\i}rez$^{\text{\eighthnote}}$ }\footnotetext{{federico.faedo[at]unimi.it, ramirez.pedro[at]mi.infn.it
}}
\bigskip

$^{\text{\quarternote}}$Dipartimento di Fisica, Universit\'a di Milano, \\
Via Celoria 16, I-20133 Milano\\

\bigskip

$^{\text{\eighthnote}}$INFN, Sezione di Milano,\\
Via Celoria 16, I-20133 Milano\\

\setcounter{footnote}{0}
\renewcommand{\thefootnote}{\arabic{footnote}}

\bigskip

\bigskip

\end{center}

\vspace*{0.1cm}

\begin{abstract}  
\begin{changemargin}{-0.95cm}{-0.95cm}
\noindent  
We derive exact relations to all orders in the $\alpha'$ expansion for the charges of a bound system of heterotic strings, solitonic 5-branes and, optionally, a Kaluza-Klein monopole. The expressions, which differ from those of the zeroth-order supergravity approximation, coincide with the values obtained when only the corrections of quadratic order in curvature are included. Our computation relies on the consistency of string theory as a quantum theory of gravity; the relations follow from the matching of the Wald entropy with the microscopic degeneracy. In the heterotic frame, the higher-curvature terms behave as delocalized sources that introduce a shift between near-horizon and asymptotic charges. On the other hand, when described in terms of lower-dimensional effective fields, the solution carries constant charges over space which coincide with those of the asymptotic heterotic fields. In addition, we describe why the Gauss-Bonnet term, which only captures a subset of the relevant corrections of quadratic order in curvature, in some cases succeeds to reproduce the correct value for the Wald entropy, while fails in others.
\end{changemargin}
\end{abstract} 

\end{titlepage}

\setcounter{tocdepth}{2}
{\small
\setlength\parskip{-0.5mm} 
\noindent\rule{15.7cm}{0.4pt}
\tableofcontents
\vspace{0.6cm}
\noindent\rule{15.7cm}{0.4pt}
}

\section{Introduction}

The fundamental objects of string theory may carry several types of charge. A well-known example is given by a D5-brane of type IIB theory wrapped on a compact manifold which, besides a unit of D5-brane charge \cite{Polchinski:1995mt},\footnote{By Dp-brane charge we mean the electric charge associated to the RR $(p+1)$-form.} carries -$\beta$ units of D1-brane charge as well, where $\beta$ is the Euler character of the wrapped space divided by 24 \cite{Bershadsky:1995qy}. The somewhat unexpected D1 charge emerges from a quantum correction, which can be read from the three point function of the RR 2-form with emission of two gravitons. The relevance of this effect can hardly be overestimated. As originally noted, the shift is necessary for consistency of string duality and the fact that left-moving ground state energy of heterotic string starts at $-1$. Moreover, the shift must be taken into account for the computation of the degeneracy; if the D5-brane is part of a bound system that can be described as a black hole, the D1-brane charge it carries is fundamental to match the microscopic degeneracy with the macroscopic entropy \cite{Sen:2007qy}. It is worth emphasizing that the D1 charge is not intrinsic to the D5-brane itself, but depends on the background on which the brane is located. Other examples of similar shifts previously noticed in the literature include \cite{Sen:1997zb, Behrndt:1998eq, Castro:2007hc, Castro:2007ci, Prester:2010cw, Grimm:2018weo}.

In this article we are interested in studying similar effects in black hole backgrounds of the heterotic theory compactified on two distinct spaces: $\mathbb{T}^4 \times \mathbb{S}^1 \times \hat{\mathbb{S}}^1$ and $\mathbb{T}^4 \times \mathbb{S}^1$. In the first option, we consider a bound state of a fundamental string (F1) wrapping $\mathbb{S}^1$ with winding number $w$ and momentum $n$, $N$ solitonic 5-branes (NS5) wrapping $\mathbb{T}^4 \times \mathbb{S}^1$ and a Kaluza-Klein monopole (KK) of charge $W$ associated with the circle $\hat{\mathbb{S}}^1$. In the second option, the configuration is identical except for the absence of a KK monopole. These are respectively known as the four- or three-charge systems. For sufficiently large $n, w, N, W$, when $g_s$ is small but non-vanishing, the gravitational interaction produces the collapse of the systems. In this regime these can be described as supersymmetric black holes with four and five non-compact dimensions in terms of classical supergravity fields to a good approximation, at least outside the event horizon. These are arguably the simplest black hole systems that can be considered in string theory. Consequently, they have been subjected to numerous studies, see \cite{Horowitz:1992jp, Sen:1994eb, Strominger:1996sh, Maldacena:1996gb} for a very limited list of references. Using type II/heterotic string duality, it is possible to compare the microscopic degeneracy of the system at vanishing string coupling computed in the former theory with the entropy associated to the horizon of the black hole solution of the latter. Being BPS, the degeneracy of the system is protected under variations of $g_s$. The precise matching of both quantities constitutes a major achievement of the theory. While the agreement was first revealed for the leading order contribution, subsequent works concluded that it extends to all orders in the $\alpha'$ expansion, see \cite{Mohaupt:2000mj, Sen:2007qy, Prester:2010cw} and references therein. In the black hole description, $\alpha'$-corrections arise in the form of higher-curvature terms added to the effective action, complicating the analysis. Nevertheless, the attractor mechanism \cite{Ferrara:1995ih, Strominger:1996kf} was cleverly exploited to decouple the near-horizon region from the rest of the spacetime and study some of its properties, including the entropy \cite{Behrndt:1998eq, LopesCardoso:1998tkj, LopesCardoso:1999cv, LopesCardoso:1999fsj, Sahoo:2006pm, Prester:2008iu}. 

While very successful for this purpose, the consideration of only near-horizon regions leaves aside relevant aspects of these systems. In recent years there has been a renewed interest in exploring this territory \cite{Cano:2018qev, Chimento:2018kop, Cano:2018brq, Cano:2018hut}. The first perturbative corrections beyond the near-horizon region have been obtained for the three- and four-charge systems. Besides the particular distortion of the field configuration, which will be subjected to further modifications order by order in the $\alpha'$ expansion, these works revealed that the charges (and mass) associated to some of the constituents of the configuration suffer a shift mediated by the higher-curvature interactions. The phenomenon has a clear interpretation: the corrections modify the equations of motion order by order, introducing delocalized sources with a non-Abelian character\footnote{Actually, this line of research was triggered by the study of non-Abelian black holes in theories of supergravity coupled to Yang-Mills fields \cite{Bellorin:2007yp, Meessen:2015enl, Meessen:2017rwm, Cano:2017qrq, Avila:2017pwi}.}. 

There are several questions that can be posed here. It is possible to study how these sources are distributed over space, why the shifts affect only some of the charges, one can try to attribute some physical interpretation to their values, and so on. But, certainly, the most interesting question is whether or not it is possible to derive their precise value when \emph{all} the $\alpha'$ corrections have been accounted for. Even though only few terms of the infinite tower of higher-curvature corrections are explicitly known, we argue here that quantum gravitational consistency of the theory requires that, in the four-charge system, the relations
\begin{eqnarray}
\label{eq:4dintro}
Q_0=N-\frac{2}{W} , \qquad Q_v=W  ,\qquad Q_{-} Q_+ =nw\left(1+\frac{2}{N W} \right) \, ,
\end{eqnarray}

\noindent
are exact in the $\alpha'$ expansion. The corresponding relations in the three-charge system are 
\begin{eqnarray}
\label{eq:3dintro}
Q_0=N-1 , \qquad Q_{-} Q_+ =nw\left(1+\frac{2}{N} \right) \, .
\end{eqnarray}

\noindent
Here $Q_0$, $Q_v$, $Q_-$ and $Q_+$ are, respectively, the asymptotic charges associated to NS5-branes, KK monopole, winding and momentum of the F1. While our considerations only impose a value for the product $Q_{-} Q_+$, it seems likely that duality arguments can be used to disentangle this expression. The relations \eqref{eq:4dintro} are already satisfied when the quadratic corrections in curvature are accounted for, with $Q_-=w$. This suggests that the origin of the shift in the charges can be found in the introduction of a Chern-Simons term in the field strength of the Kalb-Ramond 2-form, followed by its corresponding supersymmetrization in the action. Hence, the shifts at first order in $\alpha'$ would be invariant under further corrections. Actually, this is what happens with the corrections to the entropy implied by Wald's formula; despite the infinite number of higher-curvature terms expected, the Chern-Simons term is the sole responsible of the modification of the Bekenstein-Hawking leading order entropy \cite{Kraus:2005vz}. Therefore, it is possible to gain very relevant information from the first set of corrections.

The above relations follow from imposing equality of the microscopic degeneracy, expressed in terms of the charges, and Wald entropy, which we compute in terms of the number of fundamental objects. In our analysis, the exact entropy can be obtained due to the non-renormalization of the near-horizon solution. This seems to be an equivalent description of the fact that the central charges of the dual CFT can be computed from the analysis of the anomalies of the theory \cite{Kraus:2005zm}, which are fully described at first order in $\alpha'$. Our result is also consistent with non-renormalization arguments for black hole entropy in the context of four-dimensional supergravity \cite{deWit:2010za}. 

The structure of the paper goes as follows. In section \ref{sec:thsol} we briefly review the heterotic theory with all the relevant corrections of quadratic order in curvature (or first order in $\alpha'$) and the perturbative three- and four-charge black hole solutions. In section \ref{sec:entch} we compute the Wald entropy of both systems and obtain the relations for the charges previously presented. Since the near-horizon geometry of the three-charge system is identical to that of the four-charge system with unit KK monopole charge, it follows that the expressions for the Wald entropy in terms of the fundamental objects are identical for both systems by setting $W=1$. In section \ref{sec:near} we make contact with previous works in the literature that studied the near-horizon regions using lower-dimensional effective actions. We identify why the inclusion of only a partial subset of corrections, like the Gauss-Bonnet term, is unable to reproduce the relevant properties of the solution for the three-charge system \cite{Cvitan:2007en}, while it succeeds for the four-charge system \cite{Behrndt:2005he}. Section \ref{sec:discuss} contains some further discussion.

\section{$\alpha'$-corrected heterotic black holes}
\label{sec:thsol}

\subsection{The theory}
\label{sec:theory}

Heterotic string theory is effectively described at low energies as a theory of classical fields in terms of a double perturbative expansion in $\alpha'$ and $g_s$. The zeroth-order term in the expansion corresponds to $\mathcal{N}=1$ supergravity, which gives a good description for sufficiently small values of curvature and string coupling. Here we shall deal with black hole solutions of sufficiently large horizon, such that the supergravity approximation is valid. Still, we are interested in performing a precision study of the information that is lost in the truncation of the subsequent terms in the expansion, and how much of this information can be restored with the tools we have. We are interested in studying the $\alpha'$ expansion, keeping ourselves content with the tree-level effective action.

The effective action of the heterotic superstring at first order in $\alpha'$ is given by \cite{Bergshoeff:1989de}

\begin{equation}
\label{action}
{S}
=
\frac{g_{s}^{2}}{16\pi G_{N}^{(10)}}
\int d^{10}x\sqrt{|{g}|}\, 
e^{-2{\phi}}\, 
\left\{
{R} 
-4(\partial{\phi})^{2}
+\frac{1}{2\cdot 3!}{H}^{2}
-\frac{\alpha'}{8}R_{(-)}{}_{\mu\nu}{}^a{}_b R_{(-)}{}^{\mu\nu\, b}{}_a +\dots
\right\}\, .
\end{equation}
We have not included Yang-Mills fields in the theory for simplicity\footnote{Some examples with non-trivial Yang-Mills fields were given in \cite{Cano:2018qev, Chimento:2018kop, Cano:2018brq}.}. Here, $R_{(-)}{}^a{}_b$ is the curvature of the torsionful spin connection defined as $\omega_{(-)}{}^a{}_b=\omega^a{}_b-\frac{1}{2}H_\mu{}^a{}_b \, dx^\mu$, namely 
\begin{equation}
R_{(-)}{}^a{}_b = d\omega_{(-)}{}^a{}_b - \omega_{(-)}{}^a{}_c \wedge \omega_{(-)}{}^c{}_b \, .
\end{equation}

\noindent
The field strength $H$ of the Kalb-Ramond 2-form $B$ includes the Chern-Simons term 
\begin{equation}
\label{def:kalb-ramond}
H = dB+ \frac{\alpha'}{4} \Omega^{\text{L}}_{(-)} \,,
\end{equation}
where
\begin{equation}
\label{def:chern-simons}
\Omega^{\text{L}}_{(-)} = d\omega_{(-)}{}^a{}_b \wedge \omega_{(-)}{}^b{}_a - \frac{2}{3} \omega_{(-)}{}^a{}_b \wedge \omega_{(-)}{}^b{}_c \wedge \omega_{(-)}{}^c{}_a \,.
\end{equation}
The corresponding Bianchi identity reads
\begin{equation}\label{eq:bianchi}
dH=\frac{\alpha'}{4}R_{(-)}{}^a{}_b\wedge R_{(-)}{}^b{}_a \, ,
\end{equation}

\noindent
while the equations of motion are 
\begin{eqnarray}
\label{eq:eq1}
R_{\mu\nu} -2\nabla_{\mu}\partial_{\nu}\phi
+\frac{1}{4}{H}_{\mu\rho\sigma}{H}_{\nu}{}^{\rho\sigma}
-\frac{\alpha'}{4}R_{(-)}{}_{\mu\rho}{}^a{}_b R_{(-)}{}_\nu{}^{\rho\, b}{}_a
& = & 
\mathcal{O}(\alpha'^2)\, ,
\\
& & \nonumber \\
\label{eq:eq2}
(\partial \phi)^{2} -\frac{1}{2}\nabla^{2}\phi
-\frac{1}{4\cdot 3!}{H}^{2}
+\frac{\alpha'}{32}R_{(-)}{}_{\mu\nu}{}^a{}_b R_{(-)}{}^{\mu\nu\, b}{}_a
& = &
\mathcal{O}(\alpha'^2)\, ,
\\
& & \nonumber \\
\label{eq:eq3}
d\left(e^{-2\phi}\star\!{H}\right)
& = &
\mathcal{O}(\alpha'^2)\, .
\end{eqnarray}

\noindent
The zeroth-order supergravity theory can be recovered from these expressions by setting $\alpha'=0$. Moreover,  any solution to the above equations of motion satisfying $R_{(-)}{}^a{}_b =0$ is also a solution of the zeroth-order supergravity theory. This is a simple observation that plays a very important role; for the families of supersymmetric black holes that we shall consider $R_{(-)}{}^a{}_b$ vanishes in the near-horizon limit, while it is non-zero in the exterior region of the black hole. Therefore, the higher-curvature corrections do not alter the fields at the horizon, although they introduce modifications in the external region interpolating to asymptotic infinity.

Notice that the action includes a tower of corrections of all powers in $\alpha'$ due to the recursive definition of the Kalb-Ramond field strength. Actually, the term of quadratic order in curvature at \eqref{action} was found imposing supersymmetry of the theory at first order in $\alpha'$ after including the Chern-Simons term \cite{Bergshoeff:1988nn}. Further corrections of higher power in the curvature $R_{(-)}$ of the torsionful spin connection are required to recover supersymmetry order by order. The quartic effective action of heterotic theory, constructed in \cite{Bergshoeff:1989de}, was also obtained using this criterion. On the other hand, additional higher-curvature corrections unrelated to the supersymmetrization of the Kalb-Ramond kinetic term also appear. Not much is known about them, although it has been conjectured that it should be possible to write them in terms of contractions of the curvature $R_{(-)}$ and the metric. We refer to \cite{Prester:2010cw} for a description of this issue.

\subsection{Four-charge black hole}


A perturbative solution to first order in $\alpha'$ of the equations \eqref{eq:bianchi}-\eqref{eq:eq3} was found in \cite{Chimento:2018kop, Cano:2018brq}. The fields are expressed in terms of four functions $\mathcal{Z}_{\pm,0}$ and $\mathcal{V}$, 
\begin{eqnarray}
\nonumber
ds^{2}
& = &
\frac{2}{\mathcal{Z}_{-}}du
\left[dt-\frac{1}{2}\mathcal{Z}_{+} du\right]
-\mathcal{Z}_{0}d\sigma^{2}_{(4)}
-d\vec{y}^2\, ,
\\
& & \nonumber \\ \nonumber
e^{-2{\phi}}
& = &
g_s^{-2}\frac{\mathcal{Z}_{-}}{\mathcal{Z}_{0}}\, ,
\\
& & \nonumber \\
\label{eq:fieldconf}
H 
& = & 
d\mathcal{Z}^{-1}_{-}\wedge du \wedge dt+\star_{(4)}d\mathcal{Z}_{0}
\, ,  
\end{eqnarray}

\noindent
where the Hodge dual in the last equation is associated to the four-dimensional metric $d\sigma^2_{(4)}$, which is a Gibbons-Hawking (GH) space:
\begin{equation}
d\sigma^2_{(4)}=\mathcal{V}^{-1} \left( d z + \chi \right)^2+
\mathcal{V} d\vec{x}^2_{(3)}
\, , \qquad
d \mathcal{V}=\star_{(3)} d \chi \, .
\label{eq:GH}
\end{equation}

\noindent
It is further assumed that $\mathcal{Z}_{\pm,0}$ and $\mathcal{V}$ only depend on the coordinates $\vec{x}_{(3)}$ that parametrize $\mathbb{E}^3$. Before specifying a precise form for these functions, these expressions describe a field configuration preserving four supercharges whose compactification in the $u$ coordinate yields a static spacetime\footnote{To describe the most general field configuration with these properties, $d\sigma^2_{(4)}$ is taken as a generic hyperK\"ahler space on which $\mathcal{Z}_{\pm,0}$ vary.}. A spherically symmetric (in $\vec{x}_{(3)}$) solution to the equations of motion is given by
\begin{align}
\nonumber
\mathcal{V} 
& = 
1+\frac{q_v}{r} \, ,\\
\nonumber
\mathcal{Z}_{-} 
& = 
1+\frac{q_{-}}{r} \, ,\\
\nonumber
\mathcal{Z}_{0} 
& = 
1+\frac{q_{0}}{r}- \alpha'\left[F(r;q_{0})+F(r;q_v)
\right] \, ,\\
\label{eq:Zs}
\mathcal{Z}_{+} 
& = 
1+\frac{q_{+}}{r}+\frac{\alpha' q_{+}}{2q_v q_{0}} 
\frac{r^{2}+r(q_{0}+q_{-}+q_v)+q_v q_{0}+q_v q_{-}+q_{0}q_{-}
}{(r+q_v)(r+q_{0})(r+q_{-})}\, ,
\end{align}
\noindent
where  
\begin{equation}
\label{eq:Fdef}
F(r;k) 
:=\frac{(r+q_v)(r+2k)+k^{2}}{4q_v(r+q_v)(r+k)^{2}}\, .
\end{equation}

Again, one can recover the solution to the zeroth-order supergravity theory simply by setting $\alpha'=0$, obtaining four harmonic functions. The corrections to the harmonic leading terms are in all cases finite and their absolute value is monotonically decreasing. In the near-horizon limit, $r\rightarrow 0$, when the corrections take their largest absolute value, their effective contribution is actually zero. The harmonic poles of the zeroth-order solution are responsible for the existence of this well-known decoupling regime. Therefore, the near-horizon solution is unaltered by the correction. Another way to understand this important fact is to study the near-horizon solution in its own, which reads
\begin{eqnarray}
\nonumber
ds^{2}
& = &
\frac{2r}{q_{-}}du
\left[dt-\frac{q_{+}}{2r} du\right]-
q_{0} q_v \left[ \frac{dr^2}{r^2}+d\theta^2+\sin^2 \theta d\varphi^2+\left(\frac{dz}{q_v}+ \cos \theta d\varphi \right)^2 \right]
-d\vec{y}^2\, ,
\\
& & \nonumber \\ \nonumber
e^{-2{\phi}}
& = &
g_s^{-2}\frac{q_{-}}{q_{0}}\, ,
\\
& & \nonumber \\
\label{eq:nH4c}
H 
& = & 
\frac{1}{q_-} dr \wedge du \wedge dt+q_0 \sin\theta d\theta  \wedge dz \wedge d\varphi
\, .
\end{eqnarray}

\noindent
The explicit computation of the curvature of the torsionful spin connection for the near-horizon solution yields $R_{(-)}{}^a{}_b =0$. Then, as previously stated, \eqref{eq:nH4c} remains the same in the truncation to the supergravity approximation. 

The identification of the $q_i$ parameters in terms of localized, fundamental objects of string theory has been performed in \cite{Cano:2018brq}. From the preceding discussion, one sees that such relations can be obtained using the standard techniques on the near-horizon solution of the simpler supergravity theory. The result is
\begin{eqnarray}
\label{eq:qs}
q_{+}&=&\frac{\alpha'^2 g_{s}^{2}n}{ 2 R_z R_{u}^{2}}\, , \qquad q_{-}=\frac{\alpha' g_{s}^{2}w}{2 R_z }\, , \qquad 
q_0=\frac{\alpha' N}{2 R_z}\, , \qquad q_v=\frac{W R_z}{2}\, .
\end{eqnarray}

\noindent
The system describes:
\begin{itemize}
\item{a string wrapping the circle $\mathbb{S}^1$ parametrized by $u \in (0,2\pi R_u)$ with winding number $w$ and momentum $n$,}
\item{a stack of $N$ solitonic 5-branes (NS5) wrapped on $\mathbb{T}^4 \times \mathbb{S}^1$,}
\item{a Kaluza-Klein monopole (KK) of charge $W$ associated with the circle $\hat{\mathbb{S}}^1$  parametrized by $z \in (0,2\pi R_z)$.}
\end{itemize}

The constituents have four types of charge associated. While $w$ and $N$ behave, respectively, as electric and magnetic localized sources of Kalb-Ramond charge, $n$ and $W$ correspond to momentum carried along the corresponding compact circles. Additionally, the higher-curvature terms induce self-interactions that behave as delocalized charge sources. For the system studied, the non-vanishing terms responsible for this effect occur at the Bianchi identity \eqref{eq:bianchi} and the $uu$ component of the Einstein equation \eqref{eq:eq1}, which produce deviations of the functions $\mathcal{Z}_{0,+}$ from the leading harmonic term. They introduce solitonic 5-brane and string momentum charge densities distributed in the exterior of the black hole horizon. The charge contained inside a sphere of radius $r^*$ is $Q_{i,r^*} \sim r^2 \partial_r \mathcal{Z}_i \vert_{r=r^*}$. The total, asymptotic charges are\footnote{We normalize the charges such that they are independent of the moduli.}
\begin{eqnarray}
\label{eq:charges4}
Q_{+}=n+\frac{2n}{N W} , \qquad Q_{-}=w , \qquad 
Q_0=N-\frac{2}{W} , \qquad Q_v=W  .
\end{eqnarray}

The computation of the ADM mass of the black hole yields
\begin{equation}\label{M2}
M
=
\frac{1}{R_{u}}\left(n+\frac{2n}{N W}\right) 
+\frac{R_{u}}{\ell_{s}^{2}}w
+\frac{R_{u}}{g_{s}^{2}\ell_{s}^{2}}\left(N-\frac{2}{W}\right)
+\frac{R_z^{2} R_{u}}{g_{s}^{2}\ell_{s}^{4}} W \, .
\end{equation}

\noindent
Being supersymmetric and, hence, extremal, the mass of the black hole coincides with the sum (up to moduli factors) of the four charges associated to the constituents. This computation reveals that the charge-to-mass ratio of these configurations is not modified by higher-curvature corrections, a behaviour that has been argued to occur in non-supersymmetric extremal black holes \cite{Cheung:2018cwt, Bellazzini:2019xts, Aalsma:2019ryi}.

In first instance, additional higher-curvature corrections will behave as new delocalized charge sources, modifying the explicit expressions of the functions in \eqref{eq:Zs} and, presumably, the asymptotic charges $Q_i$ and ADM mass $M$. However, it was shown in \cite{Cano:2018hut} that the asymptotic solitonic 5-brane charge $Q_0$ is protected under further corrections. In section \ref{sec:entch} we review this result and obtain exact relations for the rest of the charges in the $\alpha'$ expansion.

\subsection{Three-charge black hole}

A simpler black hole solution can be described if the KK monopole is removed from the previous configuration. The field structure in \eqref{eq:fieldconf} is preserved, while the four-dimensional hyperKähler manifold is simply $\mathbb{R}^4$,

\begin{equation}
d\sigma^2_{(4)}=d\rho^2+\rho^2 d\Omega^2_{(3)} \, .
\label{eq:R4}
\end{equation}

\noindent
This particular case can also be described as a Gibbons-Hawking space with $\mathcal{V}=R_z/(2r)$, introducing a new radial variable $r=\rho^2/(2R_z)$. Then, the near-horizon geometry is identical to that of the four-charge system with $W=1$. The complete solution reads
\begin{align}
\nonumber
\mathcal{Z}_{-} 
& = 
1+\frac{\tilde{q}_{-}}{\rho^2} \, ,\\
\nonumber
\mathcal{Z}_{0} 
& = 
1+\frac{\tilde{q}_{0}}{\rho^2}- \alpha' \frac{\rho^2+2\tilde{q}_0}{(\rho^2+\tilde{q}_0)^2} \, ,\\
\label{eq:Zs3q}
\mathcal{Z}_{+} 
& = 
1+\frac{\tilde{q}_{+}}{\rho^2}+\frac{2\alpha' \tilde{q}_{+}}{\tilde{q}_{0}} 
\frac{\rho^2+\tilde{q}_0+\tilde{q}_-}{(\rho^2+\tilde{q}_0)(\rho^2+\tilde{q}_{-})}\, ,
\end{align}

\noindent
where we introduced $\tilde{q}_i=2R_z q_i$ for convenience. The near-horizon solution is

\begin{eqnarray}
\nonumber
ds^{2}
& = &
\frac{2\rho^2}{\tilde{q}_{-}}du
\left[dt-\frac{\tilde{q}_{+}}{2\rho^2} du\right]-
\tilde{q}_{0} \left[ \frac{d\rho^2}{\rho^2}+ \frac{1}{4}\left(d\theta^2+d\psi^2+d\varphi^2+2 \cos \theta d\varphi d\psi \right) \right]
-d\vec{y}^2\, ,
\\
& & \nonumber \\ \nonumber
e^{-2{\phi}}
& = &
g_s^{-2}\frac{\tilde{q}_{-}}{\tilde{q}_{0}}\, ,
\\
& & \nonumber \\
\label{eq:nH3c}
H 
& = & 
\frac{2\rho}{\tilde{q}_-} d\rho \wedge du \wedge dt+\frac{\tilde{q}_0}{4} \sin\theta d\theta  \wedge d\psi \wedge d\varphi
\, ,  
\end{eqnarray}

\noindent
with $\psi=2z/R_z$. The $\tilde{q}_i$ parameters are 

\begin{eqnarray}
\label{eq:Qs}
\tilde{q}_{+}&=&\frac{\alpha'^2 g_{s}^{2}n}{ R_{u}^{2}}\, , \qquad \tilde{q}_{-}={\alpha' g_{s}^{2}w}\, , \qquad 
\tilde{q}_0={\alpha' N}\, ,
\end{eqnarray}

\noindent
in agreement with \eqref{eq:qs}. The total, asymptotic charges are

\begin{eqnarray}
\label{eq:charges3}
Q_{+}=n+\frac{2n}{N} , \qquad Q_{-}=w , \qquad 
Q_0=N-1 \, .
\end{eqnarray}

\noindent
Likewise, the mass of the solution is of the form of \eqref{M2} after taking into consideration the expressions for the three charges of the solution \eqref{eq:charges3}.

\section{Exact entropy and charges in the $\alpha'$ expansion}
\label{sec:entch}

In this section we compute the Wald entropy of these black holes. As already mentioned, the near-horizon solution is unaltered by the addition of quadratic terms in curvature and, moreover, it is expected to be invariant under further higher-curvature corrections. Moreover, due to the presence of an $AdS$ factor in the near-horizon geometry, the Wald entropy remains unmodified beyond first order in $\alpha'$ \cite{Kraus:2005vz, Kraus:2005zm}. Then, it is possible to compare this result with $\alpha'$-exact computations of the degeneracy obtained from microscopic counting. 


\subsection{Rewriting of the action}
\label{sec:rewriting}

The presence of Chern-Simons terms in the Kalb-Ramond field strength $H$ has been recognized to hamper the direct application of Wald's entropy formula to the action. The reason is that, even if the theory is invariant under anomalous Lorentz gauge transformations, it is difficult to express the functional dependence of $H$ on the Riemann curvature tensor in a manifestly covariant manner. For this reason, following \cite{Sahoo:2006pm, Prester:2010cw} among others, it is convenient to rewrite the action in a classically equivalent manner in terms of the dual of this field strength, whose Bianchi identity is not anomalous. Such transformation involves the addition of total derivative terms which leave the entropy invariant, according to \cite{Iyer:1994ys}, and can therefore be applied for this purpose. 

In first place, we perform a (trivial) dimensional reduction of the action to six dimensions by compactifying on $\mathbb{T}^4$ and truncating all the Kaluza-Klein modes. The solutions we consider are of course consistent with this truncation. We obtain

\eq
\label{action6d}
S = \frac{g_s^2}{16\pi G_N^{(6)}} \int d^{6}x \sqrt{|g|} \, e^{-2\phi} \left\{ R - 4(\partial\phi)^2 + \frac{1}{2\cdot 3!}{H^2} - \frac{\alpha'}{8} R_{(-)}{}_{\mu\nu}{}^a{}_b R_{(-)}{}^{\mu\nu\, b}{}_a + \dots \right\} \,,
\qe
where $G_N^{(6)} = G_N^{(10)}/\text{Vol}(\mathbb{T}^4)$. We now introduce the dual 3-form field strength $\tilde{H}=d\tilde{B}$ as $\tilde{H} \equiv e^{-2\phi} \star H$, and define the equivalent Lagrangian
%
\eq
\label{eq:lagdual}
\widetilde{S} = S + \frac{g_s^2}{16\pi G_N^{(6)}} \int \left[ \tilde{H} \wedge H - \frac{\alpha'}{4}  \tilde{H} \wedge \Omega^{\text{L}}_{(-)} \right] \, ,
\qe

\noindent
in which $\tilde{B}$ is considered a fundamental field, while $H$ is now an auxiliary field. The equation of motion of $\tilde{B}$ yields 

\eq
d\left( H - \frac{\alpha'}{4} \Omega^{\text{L}}_{(-)} \right) = 0 \,,
\qe

\noindent
whose general solution is of the form \eqref{def:kalb-ramond}. On the other hand, the dual Bianchi identity $d \tilde{H}=0$ is equivalent to \eqref{eq:eq3}. It is straightforward to check that the remaining equations of motion obtained taking $H$ as an auxiliary field are identical to those derived from the original action \eqref{action6d}. 

In this form, the modified Lagrangian is manifestly covariant except for the explicit presence of the Chern-Simons 3-form in the last term of \eqref{eq:lagdual}. The next convenient step is to decompose the Chern-Simons 3-form into a standard Chern-Simons 3-form constructed from the Levi-Civita connection and an additional contribution,
\eq
\label{eq:decomCS}
\Omega^{\text{L}}_{(-)} = \Omega^{\text{L}} + \mathscr{A} \,,
\qe
where $\Omega^{\text{L}}$ is the standard Lorentz Chern-Simons term, defined as in~\eqref{def:chern-simons}, but in terms of the spin connection $\omega^a_{\ b}$, and
\eq
\begin{aligned}
\mathscr{A} = \frac12 \, d(\omega^a_{\ b} \wedge H^b_{\ a}) + \frac14 H^a_{\ b} \wedge DH^b_{\ a} - R^a_{\ b} \wedge H^b_{\ a} + 
\frac{1}{12} H^a_{\ b} \wedge H^b_{\ c} \wedge H^c_{\ a} \,,
\end{aligned}
\qe
where $H^a_{\ b} = H_{\mu\ b}^{\ a} dx^\mu$ and $D$ is the covariant derivative operator, whose action on $H^a_{\ b}$ is $DH{}^a_{\ b} = dH{}^a_{\ b} + \omega^a_{\ c} \wedge H{}^c_{\ b} - \omega^c_{\ b} \wedge H{}^a_{\ c}$. Once plugged in the action, the first term in the above expression becomes a total derivative, so it does not enter the equations of motion or the Wald entropy. Once this term is eliminated, the contribution from $\mathscr{A}$ is manifestly covariant. 

Finally, the standard Lorentz Chern-Simons term can also be written in a manifestly covariant form by exploiting the isometries of the spacetimes considered \cite{Sahoo:2006vz}. From \eqref{eq:metricdecomp} one sees that, after compactifying on $\mathbb{T}^4$, the six-dimensional spacetime can be described as the product of two three-dimensional spaces of the form

\eq
\label{eq:3dmetric}
ds^2_{(3)} =\lambda^2 \left[ {}^{(2)} \bar{g}_{mn} dx^m dx^n - \bigl(dy + \bar{A}_m dx^m\bigr)^2  \right]\, , \qquad \, \, \, m,n=0,1 \, ,
\qe

\noindent
with $(x^0, x^1, y)$ corresponding to the coordinates $(t, r, u)$ and $(\theta, \varphi, z)$, respectively. The dual 3-form $\tilde{H}$ also factorizes in these two spaces. Hence, the remaining term in the action splits in two portions

\begin{equation}
\label{eq:decom3d}
\tilde{H} \wedge \Omega^{\text{L}} = \tilde{H}_A \wedge \Omega^{\text{L}}_B  -  \Omega^{\text{L}}_A \wedge \tilde{H}_B  \, ,
\end{equation}

\noindent
where the $A,B$ indices refer to the two different three-dimensional spaces. From this point, we continue the rewriting of the action distinguishing between the two families of solutions that we consider. For the four-charge family, the periodic coordinates $u$ and $z$ parametrize paths of finite length. The Lorentz Chern-Simons 3-form of a space of the form \eqref{eq:3dmetric} can locally be written as \cite{Guralnik:2003we}\footnote{Since there are two different three-dimensional spaces, there are two copies of each of the elements $\lambda$, ${}^{(2)} \bar{g}_{mn}$, $\bar{A}_m$ and so on. In order to simplify notation we have avoided the introduction of yet another index labeling these copies.}

\eq
\Omega^{\text{L}} =\frac{\bar{\varepsilon}^{mn}}{2}\left[ {}^{(2)} \bar{R} \bar{F}_{mn} + \bar{F}_{m p} \bar{F}^{pq} \bar{F}_{q  n } - \partial_{m } ( {}^{(2)}\bar{\omega}_{n}\,^{ab} \bar{F}_{ab}) \right] dx^0 \wedge dx^1 \wedge dy \,,
\qe
where objects with a bar are associated to the metric ${}^{(2)} \bar{g}_{mn}$ and $\bar{F} = d \bar{A}$.

Once again we observe that, after dropping the last term which contributes as a total derivative, we are left with a manifestly covariant expression to which we can apply Wald's formula. When doing so, the conformal factor in front of the two-dimensional metric must be taken into account. In particular, the relation between the spacetime and auxiliary metrics, $ {}^{(2)} g_{mn}={}^{(2)} \bar{g}_{mn}\lambda^2$, implies

\eq
{}^{(2)} \bar{R} = {}^{(2)}\!R \lambda^2-2 \nabla^2 \log \lambda \, .
\qe

The treatment of the three-charge family of solutions is a bit simpler. In this case, the three-dimensional space parametrized by $(\theta, \varphi, z)$ is a 3-sphere, with the coordinate $z$ parametrizing paths of infinite length at asymptotic spatial infinity. The Lorentz Chern-Simons form of a 3-sphere identically vanishes when evaluated from its definition. Hence, the first term in expression \eqref{eq:decom3d} is just zero in the three-charge family of solutions.
Notice that the decomposition \eqref{eq:3dmetric} becomes singular asymptotically, and it cannot be used to rewrite this term of the action. 

Therefore, we see that topological properties of the asymptotic space make a difference in the explicit expression of the manifestly covariant action. This fact plays a very important role in the study of these black holes from the near-horizon solution, as described in section \ref{sec:near}.

\subsection{Wald entropy}
\label{sec:wald}

The Wald entropy formula for a $(D+1)$-dimensional theory is

\begin{equation}
\label{eq:wald}
 \mathbb{S}
=
-2\pi \int_\Sigma d^{D-1}x\sqrt{|h|}\mathcal{E}^{abcd}\epsilon_{ab}\epsilon_{cd}\,,
\end{equation}

\noindent
where $\Sigma$ is a cross-section of the horizon, $h$ is the determinant of the metric induced on $\Sigma$, $\epsilon_{ab}$ is the binormal to $\Sigma$
with normalization $\epsilon_{ab}\epsilon^{ab}=-2$ and $\mathcal{E}^{abcd}$ is the equation of motion one would obtain for the Riemann tensor $R_{abcd}$ treating it as an independent field of the theory,

\begin{equation}
\mathcal{E}^{abcd} 
= \frac{g_s^2}{16\pi G_N^{(D+1)}}
\frac{\delta \mathcal{L}}{\delta R_{abcd}}\, ,
\end{equation}

\noindent
where $\mathcal{L}$ is the Lagrangian of the theory.

When first proposed, Wald's entropy formula was meant to be evaluated at the bifurcation surface of the event horizon \cite{Wald:1993nt}, so it was only defined for non-extremal black holes. In subsequent work \cite{Jacobson:1993vj}, it was shown that the expression \eqref{eq:wald} can still be used for any cross-section of the horizon $\Sigma$, provided the surface gravity is not zero. One way to understand the origin of this condition is to notice that, in the derivation of the formula, the null Killing vector that generates the horizon $\xi^\mu$ is normalized to have unit surface gravity. This Killing vector does not appear explicitly in \eqref{eq:wald}, whose position is taken by the binormal upon the use of $\mathcal{E}_{R}^{abcd}\epsilon_{ab}\epsilon_{cd}=\mathcal{E}_{R}^{abcd}\nabla_a \xi_b \nabla_c \xi_d$. When expressed in the form of \eqref{eq:wald}, Wald's entropy formula can also be evaluated for extremal black holes.

We can apply this formula to the action of the heterotic theory directly in six dimensions, after performing a trivial compactification on $\mathbb{T}^4$. It is convenient to rewrite the metric as

\begin{align}
\label{eq:metricdecomp}
ds_{(6)}^{2}  
& = 
e^{\phi-\phi_\infty}\left[ (k/k_\infty)^{-2/3}ds_{(5)}^{2}
-(k/k_\infty)^{2} \left( du-\frac{dt}{\mathcal{Z}_{+}} \right)^{2}\right]\, ,
\end{align}

\noindent
where the lower dimensional line elements, the dilaton $\phi$ and the Kaluza-Klein scalars $k$ and $\ell$ are

\begin{equation}
\begin{array}{rcl}
ds_{(5)}^{2} 
& = &
(\ell/\ell_\infty)^{-1}\, ds_{(4)}^{2}-(\ell/\ell_\infty)^{2} \left( dz+\chi \right)^{2}\, ,
\\ & & \\
ds_{(4)}^{2} 
& = & 
e^{2U}dt^{2}-e^{-2U} \left( dr^2 + r^2 d\Omega^2_{(2)} \right)\, ,
\\
& & \\
e^{2\phi}
& = &
e^{2\phi_{\infty}}{\displaystyle\frac{\mathcal{Z}_{0}}{\mathcal{Z}_{-}}}\, ,
\hspace{1cm}
k
= 
k_{\infty} 
{\displaystyle\frac{\mathcal{Z}_{+}^{1/2}}{\mathcal{Z}_{0}^{1/4}\mathcal{Z}_{-}^{1/4}}}\, ,
\hspace{1cm}
\ell
= 
\ell_{\infty} 
{\displaystyle\frac{\mathcal{Z}_{0}^{1/6}\, \mathcal{Z}_{+}^{1/6}\,
    \mathcal{Z}_{-}^{1/6}}{\mathcal{V}^{1/2}}}\, ,
\label{eq:kkscalars}
\end{array}
\end{equation}

\noindent
with $e^{\phi_{\infty}}=g_{s}$ and 

\begin{equation}
e^{-2U} = \sqrt{\mathcal{Z}_{0}\, \mathcal{Z}_{+}\, \mathcal{Z}_{-} \mathcal{V}}\, .
\end{equation}

For a four-charge configuration, $ds_{(4)}^{2}$ is the four-dimensional metric in the Einstein frame, while $\phi$, $k$ and $\ell$ provide a parametrization of the three scalars, which are real in the solution considered. The volume form entering Wald's formula is

\begin{equation}
d^4 x \sqrt{|h|}
= d\theta d\varphi dz du
\sqrt{\mathcal{Z}_{0}\, \mathcal{Z}_{+}\, \mathcal{Z}_{-} \mathcal{V}}   r^{2} \sin{\theta} e^{2(\phi-\phi_\infty)}   \,.
\end{equation}

\noindent
In order to compute the integrand it is convenient to use flat indices. We define the vielbein

\begin{gather} 
\nonumber
 e^{0} = e^{\frac{\phi-\phi_\infty}{2}}(\frac{k}{k_\infty})^{-\frac13} (\frac{\ell}{\ell_\infty})^{-1/2} e^{U}  dt\,,\qquad 
 e^{1}=  e^{\frac{\phi-\phi_\infty}{2}}(\frac{k}{k_\infty})^{-\frac13} (\frac{\ell}{\ell_\infty})^{-1/2} e^{-U}  dr\,,\qquad  
 \\ \nonumber \\ \nonumber
 e^{2} = e^{\frac{\phi-\phi_\infty}{2}}(\frac{k}{k_\infty})^{-\frac13} (\frac{\ell}{\ell_\infty})^{-1/2} e^{-U} r d\theta\,, \qquad 
 e^{3} =  e^{\frac{\phi-\phi_\infty}{2}}(\frac{k}{k_\infty})^{-\frac13} (\frac{\ell}{\ell_\infty})^{-1/2} e^{-U}r \sin\theta d\varphi\, ,
\\
\nonumber\\ 
e^{4} =  e^{\frac{\phi-\phi_\infty}{2}}(\frac{k}{k_\infty})^{-\frac13} \frac{\ell}{\ell_\infty} \left( dz+\chi \right)\, , \qquad
e^5=  e^{\frac{\phi-\phi_\infty}{2}} \frac{k}{k_\infty}
 \left( du-\frac{dt}{\mathcal{Z}_{+}} \right)\, .
\end{gather}

\noindent
In this frame, the non-vanishing components of the binormal are $\epsilon_{01}=-\epsilon_{10}=1$.

The variation of the Lagrangian with respect to the Riemann tensor contains three non-vanishing contributions. The first one comes from the Einstein-Hilbert term in \eqref{action6d}, which amounts to

\begin{equation}
\mathcal{E}_0^{abcd}= \frac{e^{-2(\phi-\phi_\infty)}}{16\pi G_N^{(6)}}\frac{ \delta R}{ \delta R_{abcd}}= \frac{e^{-2(\phi-\phi_\infty)}}{16\pi G_N^{(6)}} \eta^{ac} \eta^{bd} \, ,
\end{equation}

\noindent
where $\eta^{ab}$ is the inverse flat metric. This term is responsible for the Bekenstein-Hawking entropy $\mathbb{S}_0=A_{\Sigma}/4G_N^{(6)}$, which for large black holes gives the leading contribution to the entropy. The two additional contributions arise from the variation of the Chern-Simons 3-form in the last term of \eqref{eq:lagdual}, each one coming from one of the two factors in the decomposition \eqref{eq:decomCS}. Notice that the last term in \eqref{action6d} gives no contribution to the entropy, since it is quadratic in the curvature of the torsionful spin connection, which vanishes at the horizon. Using the rewriting performed in the previous section, in first place we get

\begin{equation}
\mathcal{E}_1^{abcd}=\frac{e^{2\phi_\infty}}{16\pi G_N^{(6)}} \frac{ \delta }{ \delta R_{abcd}} \left(- \frac{\alpha'}{(3!)^2 4} \varepsilon^{efghjk} \tilde{H}_{efg} \mathscr{A}_{hjk} \right)= \frac{e^{-2(\phi-\phi_\infty)}}{16\pi G_N^{(6)}} \frac{\alpha'}{8}H^{abf}H_f\,^{cd} \, .
\end{equation}

\noindent
To obtain the last correction to the entropy, we notice that when $\mathcal{E}^{abcd}$ gets contracted with the binormal, the only relevant values of the flat indices $a,\dots,d$ are $0,1$. Therefore, the remaining non-vanishing contribution to the entropy comes from the second term in the decomposition \eqref{eq:decom3d}, and amounts to

\begin{equation}
\mathcal{E}_2^{abcd}=\frac{e^{2\phi_\infty}}{16\pi G_N^{(6)}} \frac{ \delta }{ \delta R_{abcd}} \left(- \frac{\alpha' \varepsilon^{\mu\nu\rho\alpha\beta\gamma} }{(3!)^2 4 \sqrt{\vert g \vert}} \tilde{H}_{\mu\nu\rho} \Omega^{\text{L}}_{\alpha\beta\gamma} \right)=\frac{e^{-2(\phi-\phi_\infty)}}{16\pi G_N^{(6)}} \frac{\alpha'}{4}H^{tru} \eta^{ac} \eta^{bd} \lambda^2 \tilde{F}_{tr}  \, ,
\end{equation}

\noindent
where $t,r,u$ are curved indices, $\lambda= e^{\frac{\phi-\phi_\infty}{2}} \frac{k}{k_\infty}=\sqrt{\frac{\mathcal{Z}_+}{\mathcal{Z}_-}}$ and $\tilde{A_t}=-1/\mathcal{Z}_{+}$.

Putting everything together, Wald's entropy is
\begin{equation}
\mathbb{S}
= 
\frac{1}{4 G_{N}^{(6)}}\int d\theta d\varphi dz du 
\sqrt{q_{0} q_{+} q_{-} q}  \sin{\theta}
 \left[1+\frac{\alpha'}{4}
   \left(-H^{01f}H_f\,^{01}+H^{tru}\lambda^2 \tilde{F}_{tr} \right) \right]\,.
\end{equation}

\noindent
 The relevant components of the Kalb-Ramond field strength, in flat and curved indices, are

\begin{equation}
H_{015}
=
-(\mathcal{Z}_{0} \mathcal{V})^{-1/2} \partial_{r}\log{\mathcal{Z}_{-}}\, , \qquad
H^{tru}
=
-(\mathcal{Z}_{0} \mathcal{V})^{-1} \partial_{r}{\mathcal{Z}_{-}}\, .
\end{equation}

\noindent
Substituting these values in the expression and integrating, 

\begin{equation}
\mathbb{S}
= 
\frac{\pi}{G_{N}^{(4)}} \sqrt{q_{0} q_{+} q_{-} q} \left(1+\frac{\alpha'}{2q_0q} \right) \, ,
\end{equation}

\noindent
with the $4$-dimensional Newton constant given by

\begin{equation}
G_{N}^{(4)}=\frac{G_{N}^{(10)}}{(2\pi R_z)(2\pi R_u)(2\pi \ell_{s})^{4}}=\frac{8\pi^{6} \alpha'{}^{4} g_{s}^{2}}{(2\pi R_z)(2\pi R_u)(2\pi \ell_{s})^{4}}\, .
\end{equation}

\noindent
Using the relation between the charge parameters $q_i$ and the number of fundamental objects in the system, we finally get

\begin{equation}
\label{eq:entropy}
\mathbb{S}
=
2\pi \sqrt{ nw NW } \left(1+\frac{2}{NW}\right)\, .
\end{equation}

The entropy of the three-charge system is obtained by setting $W=1$ in this expression since, as we previously noted, the near-horizon solution is identical to that of a four-charge black hole with unit Kaluza-Klein monopole charge.

\subsection{Corrected charges}

We have obtained an expression for the Wald entropy of these families of black holes in terms of the number of fundamental objects of the solution. The result has a clear interpretation: the Chern-Simons term, which is needed for anomaly cancellation, is the sole responsible of the increase in the entropy with respect to the Bekenstein-Hawking term. The near-horizon background remains unperturbed under the curvature corrections of quadratic order, and thus the area of the event horizon is unchanged. This is a consequence of the supersymmetric structure of the theory (and the solutions), which restricts the functional form of the corrections to objects constructed from the curvature of the torsionful spin connection (which vanishes for this background) \cite{Bergshoeff:1988nn}. 

The Wald entropy can be compared with the microscopic degeneracy of the string theory system it represents, whose value is known to all orders in the $\alpha'$ expansion. For the four-charge solution it is \cite{Sen:2007qy}

\begin{equation}
\label{eq:microentropy4c}
\mathbb{S}
=
2\pi \sqrt{Q_- Q_+  \left( Q_0 Q_v+4 \right) }\, .
\end{equation}

\noindent
Here $Q_i$ are the charges corresponding to winding ($Q_-$), momentum ($Q_+$), solitonic 5-brane ($Q_0$) and Kaluza-Klein monopole ($Q_v$). The presumed quantum gravitational consistency of string theory imposes the equality of both the macroscopic and microscopic entropies. This can be used to derive exact relations between the charges and the number of fundamental objects to all orders in $\alpha'$.

There are, of course, infinitely many alternative expressions for the charge shifts that respect the equality between the macroscopic and microscopic entropies. However, there are a series of arguments that enable us to propose a set of definite relations. We start by recalling the already known exactness of

\begin{eqnarray}
\label{eq:charges4exact1}
Q_0=N-\frac{2}{W} , \qquad \qquad Q_v=W  .
\end{eqnarray}

\noindent
The non-renormalization of the KK monopole charge, $Q_v=W$, follows from the supersymmetry of the solution; any correction to the $\mathcal{V}$ function would make the $d\sigma^2$ metric no longer hyperKähler. On the other hand, the exact NS5 charge screening, $Q_0=N-\frac{2}{W}$, was first described in \cite{Cano:2018hut}. It can be obtained by integrating the Bianchi identity, whose form is dictated by the anomaly cancellation mechanism. The shift is produced by a negative NS5 charge density carried by a gravitational $SO(4)$ instanton with instanton number $2/W$, which is delocalized over the full space. Half of the instanton charge is sourced by the KK monopole, while the other half by the stack of NS5 branes itself.

Taking this information into account, the microscopic entropy is exactly equal to the Wald entropy if the shifts in the charges induced by the higher-curvature corrections satisfy

\begin{eqnarray}
\label{eq:charges4exact2}
Q_{+} Q_- =nw\left(1+\frac{2}{N W} \right) \, .
\end{eqnarray}

\noindent
Interestingly, this already occurs at first order in $\alpha'$, see \eqref{eq:charges4}. Then, either the additional higher-curvature corrections do not introduce further charge sources, or they do it in a particular way that preserves the product. Considering that the corrections become less and less relevant order by order and that the F1 charge remains unaltered by the first correction, simplicity suggests that the expressions
\begin{eqnarray}
\label{eq:charges4exact3}
Q_{+} = n\left(1+\frac{2}{N W} \right) \, , \qquad \qquad Q_- =w \, ,
\end{eqnarray}

\noindent
are exact to all orders in $\alpha'$. While from our analysis we can only assert the validity in that respect of \eqref{eq:charges4exact2}, we would find natural that the individual relations \eqref{eq:charges4exact3} hold. It might be possible to check this guess using dualities.

With respect to the three-charge system, the microscopic entropy is \cite{Shih:2005uc, Castro:2008ys, Prester:2010cw}

\begin{equation}
\label{eq:microentropy3c}
\mathbb{S}
=
2\pi \sqrt{Q_- Q_+  \left( Q_0 +3 \right) }\, .
\end{equation}

\noindent
The application of the previous arguments gives

\begin{eqnarray}
\label{eq:charges3exact}
Q_0=N-1 , \qquad \qquad Q_{+} Q_- =nw\left(1+\frac{2}{N} \right)  ,
\end{eqnarray}

\noindent
which again is satisfied already at first order in $\alpha'$.


\section{Lower dimensional, near-horizon effective approaches}
\label{sec:near}

The study of heterotic black holes and their higher-curvature corrections has been mainly approached in the literature using two different strategies.  In the first one, developed around the early 00's, the target is to find a solution of the form $AdS \times X$, with $X$ some compact manifold, characterized by a given set of charges. It is then typically assumed that such solution describes the near-horizon limit of an extremal black hole with the same charges, and its properties are subsequently studied\footnote{In this period some early work was also performed on the study of higher-curvature corrections to global solutions, see for instance \cite{LopesCardoso:2000qm}.}. Several methods have been developed to achieve this purpose, which can be applied in the context of different effective theories of interest. An intriguing result obtained from this line of investigation is that, in some cases, it is possible to reproduce the microscopic entropy by including only a subset of the curvature corrections to the action. The Gauss-Bonnet term (GB), which is known to be one of the corrections to the lower dimensional effective theory \cite{Zwiebach:1985uq}, probably provides the most interesting example; the value of the Wald entropy obtained from its inclusion correctly reproduces the microscopic degeneracy of the four-charge system, while it fails to do so for the simpler three-charge system.

The second strategy, which has been recently developed, is the one we followed in previous sections. Starting from a complete black hole solution of the theory of supergravity, the corrections induced by higher-curvature terms are computed using the standard perturbative approach. While conceptually simple, the problem is technically involved and other strategies were usually preferred. On the other hand, the benefit of this effort is that information about the solution beyond the near-horizon limit becomes available. 

At present time there are results that have been obtained using both strategies. It is, therefore, necessary to compare them and see what can be learned from the analysis. This is the aim of this section. 

\subsection{Compactification of the supergravity theory}

From an effective four-dimensional perspective, the fields relevant for the description of such system are related to the heterotic fields as follows\footnote{In order to make the comparison with previous literature transparent, the dimensional reduction of this section uses a different parametrization of the scalars than that of section \ref{sec:wald}.}
\begin{eqnarray}
\nonumber
g_{\mu\nu}
& = &
\hat{g}_{\mu\nu} - \frac{\hat{g}_{u\mu} \hat{g}_{u\nu}}{\hat{g}_{uu}}  - \frac{\hat{g}_{z\mu} \hat{g}_{z\nu}}{\hat{g}_{zz}}  \, ,
\\
& & \nonumber \\ \nonumber
s &=&
e^{-2{{\phi}}} \sqrt{\hat{g}_{uu}\hat{g}_{zz}}
\, , \qquad
t=\sqrt{\vert \hat{g}_{uu} \vert} \, , \qquad
u=\sqrt{\vert \hat{g}_{zz} \vert} \, ,
\\
& & \nonumber \\
\label{eq:compactification}
A^{(1)}_\mu 
& = & 
-\frac{\hat{g}_{u\mu}}{2\hat{g}_{uu}} 
\, , \qquad
A^{(2)}_\mu 
=
-\frac{\hat{g}_{z\mu}}{2\hat{g}_{zz}} 
\, , \qquad
A^{(3)}_\mu 
=
\frac{\tilde{B}_{u\mu}}{2} 
\, , \qquad
A^{(4)}_\mu 
=
\frac{\tilde{B}_{z\mu}}{2} 
\, . \qquad
\end{eqnarray}

\noindent
Here $(\mu ,\nu) \in (t,r,\theta,\varphi)$, and we introduced hats to distinguish the higher-dimensional metric. It is convenient to define $A^{(3,4)}$ in terms of the dual of the Kalb-Ramond 2-form, as in this manner their field strength is closed, $ F_{(3,4)}=dA^{(3,4)}$. Using this identification, the zeroth-order supergravity theory compactified to four dimensions is

\eq
\label{action4d}
S = \frac{g_s^2}{16\pi G_N^{(4)}} \int d^{4}x \sqrt{|g|} \, s \left\{ R-a_{ij} \partial_\mu\phi^i \partial^\mu\phi^j - {t^2} F_{(1)}^2 -{u^2} F_{(2)}^2-\frac{u^2}{s^2} F_{(3)}^2- \frac{t^2}{s^2} F_{(4)}^2 \right\} \,,
\qe

\noindent
where we denote the scalars collectively as $\phi^i$, with $a_{ij}$ some functions of the scalars\footnote{In the family of near-horizon solutions that we consider the scalars are constant, so the $\sigma$-model will play no role.} and $F_{(a)}^2=F_{(a)\, \mu\nu} F_{(a)}^{\mu\nu}$. 

We are interested in finding solutions to the equations of motion derived from \eqref{action4d} describing the near-horizon region of an extremal black hole. The geometry of these is known to be of the form $AdS_2 \times S^2$. The general field configuration consistent with this isometry and the set of four independent charges we consider in this article is

\begin{eqnarray}
\nonumber
ds^{2}
& = &
v_1 \left(r^2 dt^2 - \frac{dr^2}{r^2} \right) - v_2 d\Omega^2_{(2)} \, ,
\\
& & \nonumber \\ \nonumber
s &=& u_s\, , \qquad
t = u_t \, , \qquad
u= u_u \, ,
\\
& & \nonumber \\
\label{eq:nearansatz}
F^{(1)}_{rt}
& = & 
e_1
\, , \qquad
F^{(2)}_{\theta\varphi} 
=
{P_2} \sin\theta
\, , \qquad
F^{(3)}_{rt} 
=
e_3
\, , \qquad
F^{(4)}_{\theta\varphi} 
=
{P_4} \sin\theta
\, .
\end{eqnarray}

\noindent
For the three-charge system we can take the same configuration for the fields, but it is necessary to fix $P_2=\frac{R_z}{4}$. This is equivalent to the statement that the near-horizon geometry of the three-charge system is identical to that of the four-charge system with unit KK monopole. The equations of motion in this case imply $u_u=\frac{2\sqrt{v_2}}{R_z}$.\footnote{This substitution should not be made in the function $f$ defined in \eqref{eq:deff}, as this would yield incorrect equations of motion.} In this manner, there are only three independent vectors and two independent scalars, and the cross-section of the horizon contains a 3-sphere when embedded in the heterotic theory.

The well-known attractor mechanism establishes that the parameters of the solution are fully determined in terms of the charges carried by the vectors. The magnetic and electric charges are defined in the standard manner,\footnote{The normalization constants in the definition of charges have been chosen for later convenience.}

\begin{equation}
\label{eq:defch}
P_a = \frac{1}{4\pi } \int_{S^2} d\theta d\varphi F^{(a)}_{\theta\varphi} \, , \qquad
Q_a =  \frac{1}{16\pi} \int_{S^2} d\theta d\varphi \frac{\delta}{\delta F^{(a)}_{rt}}  (\sqrt{|g|} \mathcal{L})\, .
\end{equation}

\noindent 
These integrals  can be defined not only for the near-horizon geometry, but for the full black hole solution. As consequence of the Bianchi identities $\partial_r F^{(a)}_{\theta\varphi}=0$ and the Maxwell equations $\partial_r \left[ \frac{\delta}{\delta F^{(a)}_{rt}}  (\sqrt{|g|} \mathcal{L}) \right]=0$ of the vectors, the charges are independent of the radius of the sphere on which they are computed. This implies that, from the four-dimensional effective perspective, the asymptotic and near-horizon charges of the solution coincide, even after the inclusion of higher-curvature corrections. As this behaviour is different from the one displayed by the ten-dimensional fields, one should be very cautious when interpreting lower-dimensional fields in the string theory language. We will come back to this point later. For the moment, since we do not have higher-derivative terms yet, this distinction is unnecessary.

The relations between the parameters of the near-horizon background and the charges can be determined as follows \cite{Sen:2005wa}. One first defines the function

\begin{equation}
\label{eq:deff}
f(v_1, v_2, u_i, e_a, P_a) =  \int_{S^2} d\theta d\varphi \sqrt{|g|} \mathcal{L}(v_1, v_2, u_i, e_a, P_a) \, ,
\end{equation}

\noindent
where the ansatz \eqref{eq:nearansatz} is used to evaluate the right hand side. From \eqref{eq:defch} it follows 

\begin{equation}
 \frac{ 1}{16\pi}\frac{\partial f}{\partial e_a}=Q_a \, ,
\end{equation}

\noindent
which can be used to replace $e_a$ by $Q_a$ if wanted. The solution is obtained by extremizing the function $f$,

\begin{equation}
\frac{\partial f}{\partial v_1}=0 \, , \qquad
\frac{\partial f}{\partial v_2}=0 \, , \qquad
\frac{\partial f}{\partial u_i}=0 \, . 
\end{equation}

\noindent
The black hole entropy is proportional to the Legendre transformation of $f$ evaluated on the extremum, 

\begin{equation}
\label{eq:entropyfunc}
 \mathbb{S}
=
 \frac{g_s^2}{8 G_N^{(4)}} \left(16\pi e_a Q_a - f \right) \vert_{ext.}\, .
\end{equation}

\subsection{Near-horizon solutions}

It is straightforward to apply this formalism to the compactified zeroth-order heterotic theory \eqref{action4d}. We obtain

\begin{eqnarray}
\nonumber
ds^{2}
& = &
4 P_2 Q_3 \left(r^2 dt^2 - \frac{dr^2}{r^2} -  d\Omega^2_{(2)}  \right) \, ,
\\
& & \nonumber \\  \label{eq:near4d}
s &=& \sqrt{\frac{Q_1 P_4 }{Q_3 P_2}} \, , \qquad
t = \sqrt{\frac{Q_1 }{P_4}} \, , \qquad
u=  \sqrt{\frac{Q_3 }{P_2}}  \, ,
\\
& & \nonumber \\ \nonumber
F^{(1)}_{rt}
& = & 
\sqrt{\frac{P_2 P_4 Q_3}{Q_1}}
\, , \qquad
F^{(2)}_{\theta\varphi} 
=
P_2 \sin\theta
\, , \qquad
F^{(3)}_{rt} 
=
\sqrt{\frac{P_2 P_4 Q_1}{Q_3}}
\, , \qquad
F^{(4)}_{\theta\varphi} 
=
P_4 \sin\theta
\, .
\end{eqnarray}

\noindent
We chose to scale the time coordinate such that $v_1=v_2$ to allow a straightforward comparison with previous results in the literature. Using \eqref{eq:compactification} it is possible to write the solution for the heterotic fields,

\begin{eqnarray}
\nonumber
d\hat{s}^{2}
& = &
4 P_2 Q_3 \left(r^2 dt^2 - \frac{dr^2}{r^2} - d\Omega^2_{(2)} \right)
- \frac{Q_1}{P_4} \left(du- 2\sqrt{\frac{P_2 P_4 Q_3}{Q_1}} r dt \right)^2
\\
& & \nonumber
- \frac{Q_3}{P_2} \left(dz+ 2 P_2 \cos \theta d\varphi \right)^2  \, ,
\\
& & \nonumber \\ \nonumber
e^{-2{\phi}}
& = &
\frac{P_4}{Q_3}\, ,
\\
& & \nonumber \\ \nonumber
H 
& = & 
2 \sqrt{\frac{P_2 Q_1 Q_3}{P_4}} dr \wedge du \wedge dt+2Q_3 \sin\theta d\theta  \wedge dz \wedge d\varphi
\, ,
\\
& & \nonumber \\
\label{eq:near34}
\tilde{B}
&=&
2\sqrt{\frac{P_2 P_4 Q_1}{Q_3}} r du \wedge dt- 2P_4 \cos\theta  dz \wedge d\varphi
\, .
\end{eqnarray}

\noindent
The expression coincides with the near-horizon limit of our original solutions, after rescaling the time coordinate $t \rightarrow t \sqrt{q_0 q_+ q_- q_v} $ in \eqref{eq:nH4c} and dropping the irrelevant $d\vec{y}^2$ term from the metric, with the identifications

\begin{equation}
Q_1=\frac{q_+}{2g_s^2} \, , \qquad
P_2=\frac{q_v}{2} \, , \qquad
Q_3=\frac{q_0}{2} \, , \qquad
P_4=\frac{q_-}{2g_s^2} \, .
\end{equation}

\noindent
It is important to remark that these identifications hold in the zeroth-order solution, but are modified by the $\alpha'$-corrections. As we will shortly see, the variables on the left hand side correspond to the asymptotic charges while those on the right hand side represent the number of fundamental string theory objects. It is useful to write the four-dimensional solution in terms of the latter using \eqref{eq:qs},

\begin{eqnarray}
\nonumber
ds^{2}
& = &
\frac{\alpha' NW}{4} \left(r^2 dt^2 - \frac{dr^2}{r^2} -  d\Omega^2_{(2)}  \right) \, ,
\\
& & \nonumber \\  \label{eq:near4d}
s &=&\frac{\alpha'}{R_z R_u} \sqrt{\frac{nw }{NW}} \, , \qquad
t =\frac{ \sqrt{\alpha'}}{R_u} \sqrt{\frac{n }{w}} \, , \qquad
u=\frac{ \sqrt{\alpha'}}{R_z}  \sqrt{\frac{N }{W}}  \, ,
\\
& & \nonumber \\ \nonumber
F^{(1)}_{rt}
& = & 
\frac{R_u}{4}\sqrt{\frac{wNW}{n}}
\, , \, \, \, \, 
F^{(2)}_{\theta\varphi} 
=
\frac{R_z}{4}  W \sin\theta
\, , \, \, \, \, 
F^{(3)}_{rt} 
=
\frac{\alpha'}{4R_u}\sqrt{\frac{nwW}{N}}
\, , \, \, \, \, 
F^{(4)}_{\theta\varphi} 
=
\frac{\alpha'}{4 R_z} w \sin\theta
\, .
\end{eqnarray} 

\noindent
Likewise, the black hole entropy computed from \eqref{eq:entropyfunc} gives

\begin{equation}
 \mathbb{S}_0
=
 2\pi\sqrt{nwNW} \, ,
\end{equation}

\noindent
which agrees with the leading order result we obtained in the previous section.

We have obtained these expressions from the zeroth-order supergravity theory. We recall that this field configuration describes both the three- and four-charge systems, with the former being recovered simply by setting $W=1$ or $P_2=R_z/4$. As we have already stated, the higher-curvature corrections vanish for this background and leave \eqref{eq:near4d} invariant. This means that after adding all relevant higher-curvature terms to the action \eqref{action4d} arising from the compactification of \eqref{action}, the form of the function $f$ will change, but it will have an extremum at the same point in this parameter space. On the other hand, if only a subset of the corrections are implemented the corresponding solution, if exists, will typically take a different expression.

Taking into account this information, it is simple to apply the entropy function formalism to the action that includes all relevant four-derivative terms. In order to do so, it is first necessary to write the action in a manifestly covariant form, see \cite{Sen:2007qy}, as we did in section \ref{sec:rewriting}. After few lines of computation, one can check that \eqref{eq:near4d} still gives an extremum for the corrected function $f$. On the other hand, the charges carried by the four-dimensional effective fields as defined in \eqref{eq:defch} are now for the four-charge system (for simplicity we set $R_u=R_z=\sqrt{\alpha'}=4$)

\begin{equation}
\label{eq:4charges4d}
Q_1=n \left(1+\frac{2}{NW} \right) \, , \qquad
P_2=W \, , \qquad
Q_3=N-\frac{2}{W} \, , \qquad
P_4=w \, ,
\end{equation}

\noindent
while for the three-charge system these are

\begin{equation}
\label{eq:3charges4d}
Q_1=n \left(1+\frac{2}{N} \right) \, , \qquad
Q_3=N-1 \, , \qquad
P_4=w \, .
\end{equation}

\noindent
Hence, we see that the lower-dimensional vector fields carry the asymptotic charges of our original solution of the heterotic theory. It is certainly remarkable how the shift in the charges, which is mediated by the higher-curvature corrections, distinguishes between the four- and three-charge systems, even though their near-horizon background is identical. This is caused by the explicit difference in the expression of the action in both systems when written in a manifestly covariant manner, as described in section \ref{sec:rewriting}. The asymptotic structure of the systems is responsible for the effect and, therefore, it is determinant for the analysis of the near-horizon solution.

The Wald entropy is 

\begin{equation}
 \mathbb{S}
=
 2\pi\sqrt{nwNW} \left(1+\frac{2}{NW} \right) \, 
\end{equation}

\noindent
for the the four-charge system, while the expression for the three-charge system is recovered simply setting $W=1$. Naturally, the result coincides with \eqref{eq:entropy}, which provides a consistency check between the two approaches.

In most of the preceding literature, the expressions for the lower-dimensional fields and the Wald entropy are customarily given in terms of the charges carried by the vectors. After a few lines of algebraic computation, we may write for the four-charge $\alpha'$-corrected solution

\begin{eqnarray}
\nonumber
ds^{2}
& = &
4(P_2 Q_3+2) \left(r^2 dt^2 - \frac{dr^2}{r^2} -  d\Omega^2_{(2)}  \right) \, ,
\\
& & \nonumber \\  \label{eq:near4dQ}
s &=& \sqrt{\frac{P_4 Q_1 }{P_2 Q_3+4}} \, , \, \, \, \,
t = \sqrt{\frac{Q_1 (P_2 Q_3+2)}{P_4 (P_2 Q_3+4)}} \, , \, \, \, \, 
u= \sqrt{\frac{Q_3 }{P_2} \left(1+\frac{2}{P_2 Q_3} \right)}  \, , \, \, \, \, 
\\
& & \nonumber \\ \nonumber
F^{(1)}_{rt}
& = & 
\sqrt{\frac{P_4 (P_2 Q_3+4)}{Q_1}}
\, , \, \, \, \, 
F^{(2)}_{\theta\varphi} 
=
P_2 \sin\theta
\, , \, \, \, \, 
F^{(3)}_{rt} 
=
P_2  \sqrt{\frac{P_4 Q_1 }{P_2 Q_3+4}}
\, , \, \, \, \, 
F^{(4)}_{\theta\varphi} 
=
 P_4 \sin\theta
\, , 
\\
& & \nonumber \\ \nonumber
\mathbb{S}
 & = &
 2\pi\sqrt{P_4 Q_1 \left(P_2 Q_3+4 \right) } \, ,
\end{eqnarray} 

\noindent
while for the three-charge system

\begin{eqnarray}
\nonumber
ds^{2}
& = &
4(Q_3+1) \left(r^2 dt^2 - \frac{dr^2}{r^2} -  d\Omega^2_{(2)}  \right) \, ,
\\
& & \nonumber \\  
s &=& \sqrt{\frac{P_4 Q_1 }{Q_3+3}} \, , \, \, \, \, \, \, \, 
t = \sqrt{\frac{Q_1 (Q_3+1)}{P_4 (Q_3+3)}} \, , \, \, \, \,  \, \, \, 
u= \sqrt{Q_3+1}  \, , \, \, \, \, \, \, 
\\
& & \nonumber \\ \nonumber
F^{(1)}_{rt}
& = & 
\sqrt{\frac{P_4 (Q_3+3)}{Q_1}}
\, , \, \, \, \, 
F^{(2)}_{\theta\varphi} 
=
 \sin\theta
\, , \, \, \, \, 
F^{(3)}_{rt} 
=
\sqrt{\frac{P_4 Q_1 }{Q_3+3}}
\, , \, \, \, \, 
F^{(4)}_{\theta\varphi} 
=
 P_4 \sin\theta
\, , 
\\
& & \nonumber \\ \nonumber
\mathbb{S}
 & = &
 2\pi\sqrt{P_4 Q_1 \left( Q_3+3 \right) } \, .
\end{eqnarray} 

We find perfect agreement between these expressions and the results of \cite{Cvitan:2007en, Prester:2010cw}, which consider the same action as we do. As far as they can be compared, these solutions are identical to those obtained from four-dimensional $\mathcal{N}=2$ supersymmetric theories with corrections of quadratic order in curvature in terms of the Weyl tensor \cite{deWit:1996gjy, Behrndt:1998eq, LopesCardoso:1998tkj, LopesCardoso:1999cv, LopesCardoso:1999fsj, Mohaupt:2000mj}.

\subsection{The Gauss-Bonnet correction}

A particular higher-derivative correction to the effective tree level heterotic supergravity theory in four dimensions can be written in terms of the Gauss-Bonnet (GB) density \cite{Zwiebach:1985uq},

\begin{equation}
\label{eq:GBterm}
\mathcal{L}_{GB}= 2 s \left(R_{\mu\nu\rho\sigma}R^{\mu\nu\rho\sigma}-4R_{\mu\nu}R^{\mu\nu}+R^2 \right) \, .
\end{equation}

\noindent
Even though such term only represents a subset of the relevant corrections at the four-derivative level, it has been noted in the literature that its inclusion leads to the correct value of the Wald entropy in some (but not all) cases. Particularly puzzling is the fact that it seems to give the right answer for the four-charge system, while it fails for the three-charge system. We shall now reanalyze the problem here and find the origin of this behaviour.

Let us begin with the four-charge system. Using the entropy function formalism, it is possible to obtain the near-horizon solution to the GB modified theory,

\begin{eqnarray}
\nonumber
ds^{2}
& = &
4(P_2 Q_3+2) \left(r^2 dt^2 - \frac{dr^2}{r^2} -  d\Omega^2_{(2)}  \right) \, ,
\\
& & \nonumber \\   \label{eq:near4dGB}
s &=& \sqrt{\frac{P_4 Q_1}{P_2 Q_3+4}} \, , \, \, \, \, \, \, \, 
t = \sqrt{\frac{Q_1}{P_4}} \, , \, \, \, \,  \, \, \, 
u= \sqrt{\frac{Q_3}{P_2}}  \, , \, \, \, \, \, \, 
\\
& & \nonumber \\ \nonumber
F^{(1)}_{rt}
& = & 
\sqrt{\frac{P_4 (P_2 Q_3+4)}{Q_1}}
\, , \, \, \, \, 
F^{(2)}_{\theta\varphi} 
=
 P_2 \sin\theta
\, , \, \, \, \, 
F^{(3)}_{rt} 
=
P_2\sqrt{\frac{ P_4 Q_1 }{P_2 Q_3+4}}
\, , \, \, \, \, 
F^{(4)}_{\theta\varphi} 
=
 P_4 \sin\theta
\, , 
\\
& & \nonumber \\ \nonumber
\mathbb{S}
 & = &
 2\pi\sqrt{P_4 Q_1 \left( P_2 Q_3+4 \right) } \, .
\end{eqnarray}

\noindent
This solution was first derived in \cite{Sen:2005iz}. The action complemented with \eqref{eq:GBterm} is no longer supersymmetric. It corresponds to an inconsistent truncation of the bosonic sector of the heterotic theory presented in section \ref{sec:theory}. Hence, one should be cautious when interpreting \eqref{eq:near4dGB} in string theory language. Having this in mind, it seems reasonable to identify the charges of both schemes. Direct comparison with \eqref{eq:near4dQ} reveals that the GB term suffices to capture the corrections to the metric, dilaton, vectors and Wald entropy when written in terms of the charges, while it fails with the scalars $t$ and $u$. Using \eqref{eq:4charges4d}, which in this section can be interpreted as a redefinition of the parameters describing the fields, we get

\begin{eqnarray}
\nonumber
ds^{2}
& = &
4NW \left(r^2 dt^2 - \frac{dr^2}{r^2} -  d\Omega^2_{(2)}  \right) \, ,
\\
& & \nonumber \\  
s &=& \sqrt{\frac{nw}{NW}} \, , \, \, \, \, \, \, \, 
t = \sqrt{\frac{n}{w}\left(1+\frac{2}{NW}\right)} \, , \, \, \, \,  \, \, \, 
u= \sqrt{\frac{1}{W}\left(N-\frac{2}{W}\right)}  \, , \, \, \, \, \, \, 
\\
& & \nonumber \\ \nonumber
F^{(1)}_{rt}
& = & 
\sqrt{\frac{wNW}{n}}
\, , \, \, \, \, 
F^{(2)}_{\theta\varphi} 
=
 W \sin\theta
\, , \, \, \, \, 
F^{(3)}_{rt} 
=
\sqrt{\frac{nwW}{N}}
\, , \, \, \, \, 
F^{(4)}_{\theta\varphi} 
=
 w \sin\theta
\, , 
\\
& & \nonumber \\ \nonumber
\mathbb{S}
 & = &
 2\pi\sqrt{nwNW} \left(1+\frac{2}{NW}\right) \, ,
\end{eqnarray}

\noindent
which reproduces the results derived from the heterotic theory, except for the expressions of the scalars $t$ and $u$. It is useful to write the solution in terms of these variables, as it facilitates making contact with the zeroth-order solution \eqref{eq:near4d} (we still set $R_u=R_z=\sqrt{\alpha'}=4$ for simplicity here). 

We now turn our attention to the three-charge system. In preceding sections, we described that the corresponding near-horizon solution is obtained setting $W= 1$ in the expressions for the fields and using \eqref{eq:3charges4d} for the shift in the charges. In order to obtain the correct expression for the shift, it was crucial that the higher-curvature corrections to the action are different from those of the four-charge system, as a consequence of the asymptotic structure of the solutions. From this, it is obvious that the Gauss-Bonnet term will not be able to reproduce correctly the properties of the three-charge system. The GB correction has the same impact on the three- and four-charge systems. This means that it gives the \emph{right} value for the Wald entropy in both cases when expressed in terms of the number of fundamental objects, but it is unable to produce the two different shifts for the charges. Since it gives the shift compatible with the four-charge system, when expressed in terms of the charges the Wald entropy only matches in this case. Therefore, we see that the relevant aspect to understand the puzzling behaviour of the Gauss-Bonnet correction relies on its (in)ability to reproduce the right shift in the charges. 

In this sense, the GB term is of course not unique nor special. Examples of alternative corrections that produce the exact same effect in the field configuration and its properties are

\begin{equation}
\label{eq:GBalt}
\Delta \mathcal{L}= 2s \left(R_{\mu\nu\rho\sigma}R^{\mu\nu\rho\sigma}-4R_{\mu\nu}R^{\mu\nu} \right) \, , \, \qquad
\Delta \mathcal{L}= -4 s R_{\mu\nu}R^{\mu\nu}\, ,
\end{equation}

\noindent
which correspond to an even lower subset of the corrections than those provided by the GB density. The reason is that the near-horizon background is very symmetric, so the non-vanishing components of the Riemann tensor are proportional to the metric. In flat indices and for a metric of the form \eqref{eq:nearansatz} with $v_1=v_2$, 

\begin{equation}
R_{abcd}=\frac{1}{v_1} \{ -2 \eta_{a[c} \eta_{d]b} , 2 \eta_{a[c} \eta_{d]b} \} \, ,
\end{equation}

\noindent
where the two terms correspond to the $AdS_2$ and $S^2$ factors. Hence, any scalar constructed from contractions of two Riemann tensors evaluated in the near-horizon background equals $h/v_1^2$, for some number $h$. Once multiplied by $\sqrt{\vert g \vert}$, such correction is \emph{topological}, in the sense that it is independent of the metric.

\section{Discussion}
\label{sec:discuss}

The fact that an isolated KK monopole of unit charge (i.e. $W=1$) carries $-1$ unit of NS5-brane charge in heterotic theory has long been known \cite{Sen:1997zb}. As originally argued, the gravitational instanton number acts as a negative source of magnetic charge for the Kalb-Ramond field strength. This played a crucial role in testing S-duality of heterotic theory compactified on a torus. Likewise, it is understood that for a collection of unit charged separated KK monopoles, each of them contributes $-1$ unit to the NS5 charge \cite{Sen:1997js}. Again, the value is given by the negative gravitational instanton number. A single KK monopole of charge $W$, which is the configuration of interest in four-charge black holes, has gravitational instanton number\footnote{Fractional instanton numbers are relatively common, see for example \cite{PhysRevD.21.2285, Etesi:2002cc}.} $1/W$ and hence contributes negatively to the NS5 charge by this amount. The fractional value is a direct consequence of the normalization of the Chern-Simons term entering the field strength. In all the situations mentioned, the shift is obtained from

\begin{equation}
\frac{1}{16\pi^2} \int R_{(-)}\,^a\,_b \wedge R_{(-)}\,^b\,_a \, ,
\end{equation}

\noindent
which corresponds to the integral of the right hand side of the Bianchi identity.

Moreover, the presence of torsion in the spin connection has consequences in this respect. As described in \cite{Chimento:2018kop}, an additional gravitational instanton is sourced by the stack of NS5 branes. This implies that the total shift in the NS5 charge is $-2/W$ (or simply $-1$ in the absence of KK monopole). Using this information and the computation of Wald entropy, we have obtained an exact relation for the product of the total winding and momentum charges. The analysis suggests that the introduction of the Chern-Simons term and its supersymmetrization is the sole responsible of the shifts, which would imply that the relations \eqref{eq:charges4} and \eqref{eq:charges3} at first order in $\alpha'$ are actually exact.

It is somewhat surprising that, except for the shift induced by the unit charge KK monopole, such effects had remained unnoticed until quite recently, since the four-charge black hole has been largely considered in the literature. The reason seems to be that the microscopic counting is usually done in the dual Type II description, while macroscopically the near-horizon approach in lower dimensions works directly in terms of the charges, as described in section \ref{sec:near}. It should be noticed, however, that the distinction between charges and fundamental objects is crucial in the characterization of a string theory system. The interpretation of lower-dimensional effective fields in terms of string theory is, therefore, rather subtle. A significant example of this is found for the black holes with $Q_0=0$ and $NW \neq 0$, \cite{Cano:2018hut}, which were thought to provide a regularization of the singular horizon of small black holes (that do not contain NS5 nor KK) via higher-curvature corrections \cite{Dabholkar:2004yr, Dabholkar:2004dq, Sen:2004dp}. As described in \cite{Cano:2018hut}, this interpretation was based on a misidentification of the fundamental stringy objects of the solution.

In order to compute the Wald entropy, we have rewritten the action in terms of the dual of the Kalb-Ramond form, which allows to eliminate the redundancy problem in the functional dependence of $H$ on the Riemann tensor. In view of the simplicity of the result, it seems very likely that Wald's formula can also be successfully applied to the action written in terms of $H$, as  in \eqref{action}. This was attempted in \cite{Cano:2018qev, Cano:2018brq}, obtaining a correction to the Bekenstein-Hawking term that accounts for half of the total value we obtained in \eqref{eq:entropy}. In these articles, the correction was interpreted as the first term of an infinite series expansion of $\sqrt{1+\frac{2}{NW}}$ for large $NW$, following what had been done in \cite{Sahoo:2006pm} after the shift in the NS5 charge is considered. The results presented here show that such interpretation is not correct, and that it should be possible to obtain the exact result for the entropy using directly the original form of the action \eqref{action}. Yet another alternative approach to obtain Wald entropy for the heterotic theory has been recently proposed \cite{Edelstein:2019wzg}. It would be interesting to apply this formalism to these solutions.

The extension of the analysis presented here to more general dyonic black hole solutions is an interesting line of future research. The tools developed in \cite{Ortin:2019sex} will certainly be useful for that purpose. 





\section*{Acknowledgments}

We are grateful to P. Cano, G. Lopes Cardoso, A. Ruip\'erez and T. Ortin for encouraging conversations and comments on the draft. 
This work has been supported in part by the INFN. PFR would like thank the Albert Einstein Institute at Potsdam for hospitality while this work was being completed.

\renewcommand{\leftmark}{\MakeUppercase{Bibliography}}
\phantomsection
\bibliographystyle{JHEP}
\bibliography{references}
\label{biblio}

\end{document}